\documentclass[letter]{aa}
\usepackage{float}
\usepackage{graphicx}
\usepackage{natbib}
\usepackage{txfonts}
\usepackage{color}

\begin{document}

   \title{Hunting young stars in the Galactic centre. Hundreds of thousands of solar masses of young stars in the Sagittarius\,C region}
   \titlerunning{Hundreds of thousands of solar masses of young stars in the Sagittarius\,C region}
  \author{F. Nogueras-Lara
          \inst{1}                        
          }

   \institute{
    European Southern Observatory, Karl-Schwarzschild-Strasse 2, D-85748 Garching bei M\"unchen, Germany
              \email{francisco.nogueraslara@eso.org}                                    
       }
   \date{}


  \abstract
   {The Galactic centre stands out as the most prolific star-forming environment of the Galaxy when averaged over volume. In the last 30 million years, it has witnessed the formation of $\sim10^6\,M_\odot$ of stars. However, crowding and high extinction hamper their detection and, up to now, only a small fraction of the expected mass of young stars has been identified.}
   {We aim to detect hidden young stars at the Galactic centre by analysing the stellar population in Sagittarius (Sgr)\,C. This is a region at the western edge of the nuclear stellar disc whose HII emission makes it a perfect candidate to host young stars.} 
    {We built dereddened luminosity functions for Sgr\,C and a control field in the central region of the nuclear stellar disc, and fitted them with a linear combination of theoretical models to analyse their stellar population.}
   {We find that Sgr\,C hosts several $10^5\,M_\odot$ of young stars. We compared our results with the recently discovered young stellar population in Sgr\,B1, which is situated at the opposite edge of the nuclear stellar disc. We estimated that the Sgr\,C young stars are $\sim20$\,Myr old, and likely show the next evolutionary step of the slightly younger stars in Sgr\,B1. Our findings contribute to addressing the discrepancy between the expected and the detected number of young stars in the Galactic centre, and shed light on their evolution in this extreme environment. As a secondary result, we find an intermediate-age stellar population in Sgr\,C ($\sim50$\,\% of its stellar mass with an age of between 2 and 7\,Gyr), which is not present in the innermost regions of the nuclear stellar disc (dominated by stars >7\,Gyr). This supports the existence of an age gradient and favours an inside-out formation of the nuclear stellar disc.}  
   {}


   \maketitle
%

\section{Introduction}

At only $\sim8$\,kpc from Earth, the centre of the Milky Way is the closest galaxy nucleus and the only one where we can resolve stars down to milliparsec scales. It is roughly outlined by the nuclear stellar disc (NSD), a flat stellar structure of $\sim10^9\,M_\odot$ \citep[e.g.][]{Launhardt:2002nx} with a scale length of $\sim100$\,pc and a scale height of $\sim40$\,pc \citep[e.g.][]{gallego-cano2019,Sormani:2022wv}.

Despite occupying less than 1\,\% of the  volume of the Galactic disc, the NSD is responsible for up to 10\,\% of the star forming activity of  the entire Milky Way over the past $\sim100$\,Myr \citep[e.g.][]{Mezger:1996uq,Mauerhan:2010jc,Matsunaga:2011uq,Crocker:2011kx,Nogueras-Lara:2019ad,Nogueras-Lara:2022ua}. The detection of three classical Cepheids \citep{Matsunaga:2011uq} and the analysis of luminosity functions \citep[e.g.][]{Nogueras-Lara:2019ad} revealed that on the order of $\sim10^6\,M_\odot$ of stars formed there in the last 30\,Myr. Nevertheless, the known young clusters (Arches and Quintuplet) and the young stars in the region only account for a small fraction of the expected young stellar mass \citep[e.g.][]{Figer:1999fk,Figer:1999uq,Clarkson:2012fk,Clark:2021up}. This stark difference is known as the `missing clusters problem'.

The missing clusters problem is mainly due to the high crowding and stellar density in this region, in combination with the rapid dissolution of even massive clusters in the extreme environment there \citep[$\sim6$\,Myr, for further details see][]{Kruijssen:2014aa}. Moreover, the extreme extinction mostly limits the analysis of the NSD stars in the near-infrared \citep[NIR; e.g.][]{Nishiyama:2006tx,Nishiyama:2008qa,Fritz:2011fk,Nogueras-Lara:2018aa,Nogueras-Lara:2020aa,Sanders:2022wa}, hampering the photometric detection of young stars. On the other hand, a spectroscopic search for young stars is not possible given the extremely high number of sources. An alternative way of detecting young stellar associations involves analysing stellar proper motions in order to identify comoving groups with a probable common and recent origin \citep[e.g.][]{Shahzamanian:2019aa,Martinez-Arranz:2023aa}. 

\begin{figure*}
              \begin{center}
   \includegraphics[width=0.925\linewidth]{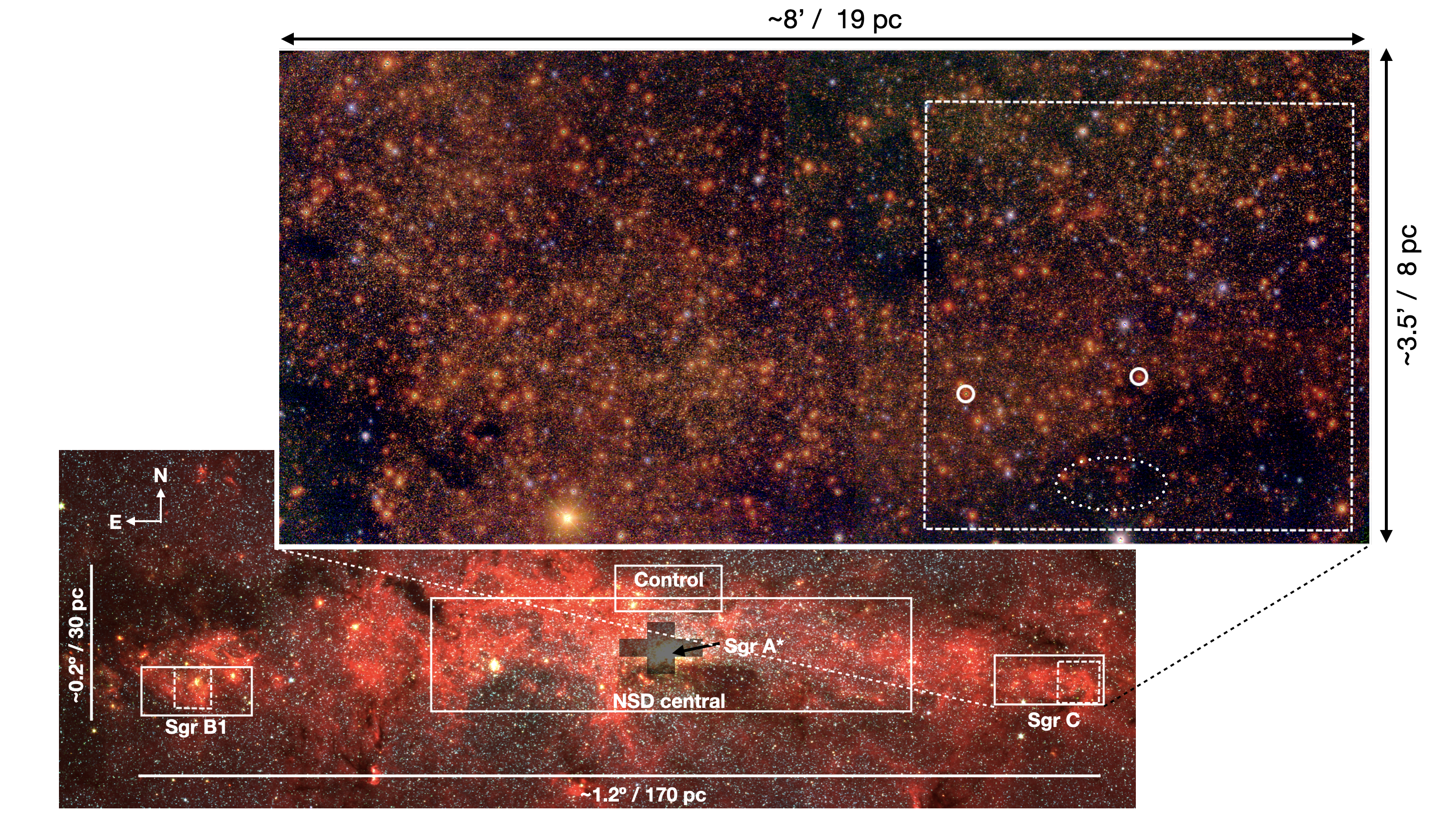}
   \caption{Spitzer false-colour image using 3.6, 4.5, and 8\,$\mu$m, as blue, green, and red, respectively \citep{Stolovy:2006fk}. The white boxes indicate the position of Sgr\,C and Sgr\,B1, as well as the control field and the central NSD region. The black shaded area shows an avoided region in the analysis of the central NSD region in \citet{Nogueras-Lara:2019ad}, corresponding to the nuclear star cluster. The small dashed rectangles indicate regions of particular interest within the Sgr\,C and Sgr\,B1 fields. The zoomed-in image corresponds to a GALACTICNUCLEUS $JHK_s$ false-colour image of the Sgr\,C field (see Sect.\,\ref{small} for details). The white circles correspond to two WCL stars \citep{Clark:2021up} and the dotted ellipse outlines a region where $H_2CO$ and $CH_3OH$ masers have been identified \citep{Caswell:1996aa,Lu:2019aa}. The compass indicates Galactic coordinates.}

   \label{scheme}
   \end{center}
    \end{figure*}

The recent analysis of Sagittarius (Sgr) B1, an HII region located at the eastern edge of the NSD \citep[e.g.][]{Simpson:2018aa}, revealed the presence of $\sim10^5\,M_\odot$ of young stars, suggesting that a significant fraction of the missing young stellar mass might be concealed within similar regions \citep{Nogueras-Lara:2022ua}. In this context, Sgr\,C emerges as a promising candidate to host a substantial number of young stars. Located at the western edge of the NSD (Fig.\,\ref{scheme}), Sgr\,C shares similarities with Sgr\,B1, including strong HII emission \citep[e.g.][]{Lang:2010aa}, ongoing star formation, and the presence of some known young stars \citep[e.g.][]{Liszt:1995aa,Forster:2000aa,Kendrew:2013aa,Lu:2019aa,Lu:2019ab,Hankins:2020aa}, and therefore represents a unique region to search for young stars and assess whether or not they follow a symmetric distribution in the NSD, considering its position relative to Sgr\,B1.

In this Letter, we present a photometric analysis of the stellar population in a $\sim8'\times 3.5'$ field covering part of the Sgr\,C region. We analysed its dereddened $K_s$ luminosity function and find that Sgr\,C contains several $10^5\,M_\odot$ of young stars, which probably belong to young dissolved clusters and stellar associations. To best of our knowledge, our analysis constitutes the first stellar population characterisation of a western region of the NSD, and reveals the presence of a significant mass of young stars there.

\section{Data}

We used $HK_s$ photometry from the GALACTICNUCLEUS survey \citep{Nogueras-Lara:2018aa,Nogueras-Lara:2019aa}. This is a high-angular resolution ($\sim0.2''$) NIR catalogue specially designed to observe the NSD. The GALACTICNUCLEUS survey reaches 5\,$\sigma$ detections at $H,K_s\sim21$\,mag, and the statistical uncertainties are below 0.05\,mag at $H\sim19$\,mag and $K_s\sim19$\,mag. We chose a field covering the Sgr\,C region \citep[D19; see Table\,A3 in][]{Nogueras-Lara:2019aa}, and also a control field observed under similar conditions at the central region of the NSD \citep[F19; see Table\,A1 in ][]{Nogueras-Lara:2019aa}. Figure\,\ref{scheme} indicates the positions of these two fields. 

The GALACTICNUCLEUS survey suffers from saturation in $K_s$ band for stars brighter than $11.5$\,mag. To avoid this problem, we replaced the photometry of saturated sources and included non-detected bright stars using the SIRIUS IRSF survey \citep[e.g.][]{Nagayama:2003fk,Nishiyama:2006tx}, as explained in \cite{Nogueras-Lara:2019ad}.

\section{Stellar population analysis}

To analyse the stellar population in Sgr\,C and the control field, we built de-reddened $K_s$ luminosity functions and fitted them with a linear combination of theoretical models applying the methodology outlined in, for example, \citet{Nogueras-Lara:2019ad,Nogueras-Lara:2022ua} and \citet{Schodel:2023aa}.

\subsection{Colour--magnitude diagram}

\begin{figure}
   \includegraphics[width=\linewidth]{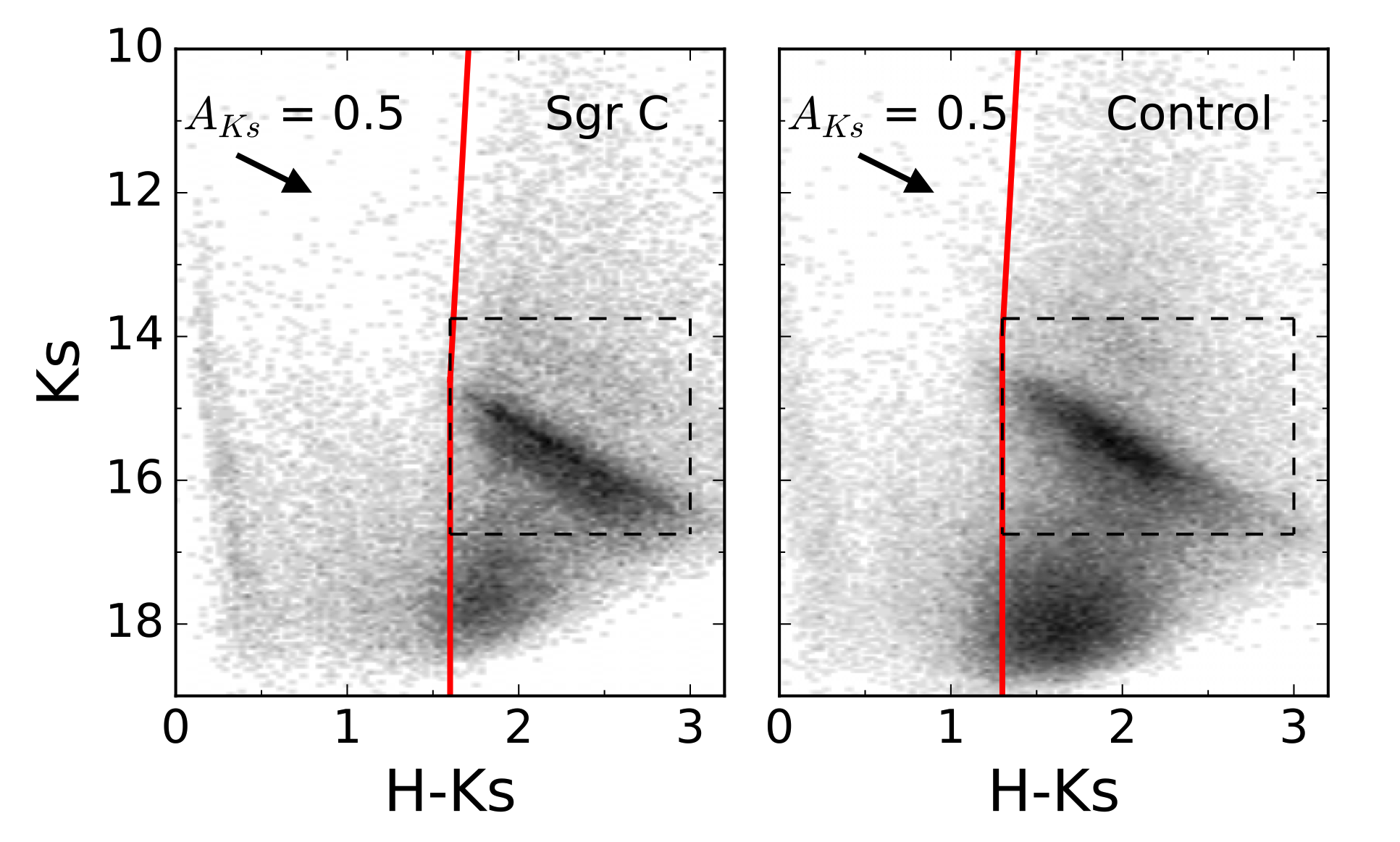}
   \caption{Colour--magnitude diagram $K_s$ versus $H-K_s$ for the Sgr\,C field (left panel) and the control region (right panel). The red lines denote the colour cuts applied to remove foreground stars. The dashed rectangles show the red giant stars used to compute the extinction maps. The black arrows indicate the direction of the reddening vector.}

   \label{CMD}
    \end{figure}

Figure\,\ref{CMD} shows the colour--magnitude diagram $K_s$ versus $H-K_s$ for the Sgr\,C region and the control field. To remove the contamination from foreground stars belonging to the Galactic disc and bar, we applied a colour cut given the significantly different extinction of these components in comparison to the Galactic centre \citep[e.g.][]{Sormani:2020aa,Nogueras-Lara:2021uz,Nogueras-Lara:2021wj}. We used red clump (RC) stars \citep[red giants in their helium core-burning sequence;][]{Girardi:2016fk}, as a reference for the colour cut, because they appear as a clear over-density in the colour--magnitude diagram and indicate the colour at which Galactic centre stars dominate. We chose $H-K_s\sim1.6$\,mag and $H-K_s\sim1.3$\,mag for the Sgr\,C and the control field, respectively. The redder cut for the Sgr\,C region indicates that the extinction is larger for this field.

\subsection{Extinction maps}
    
\begin{figure}
   \includegraphics[width=\linewidth]{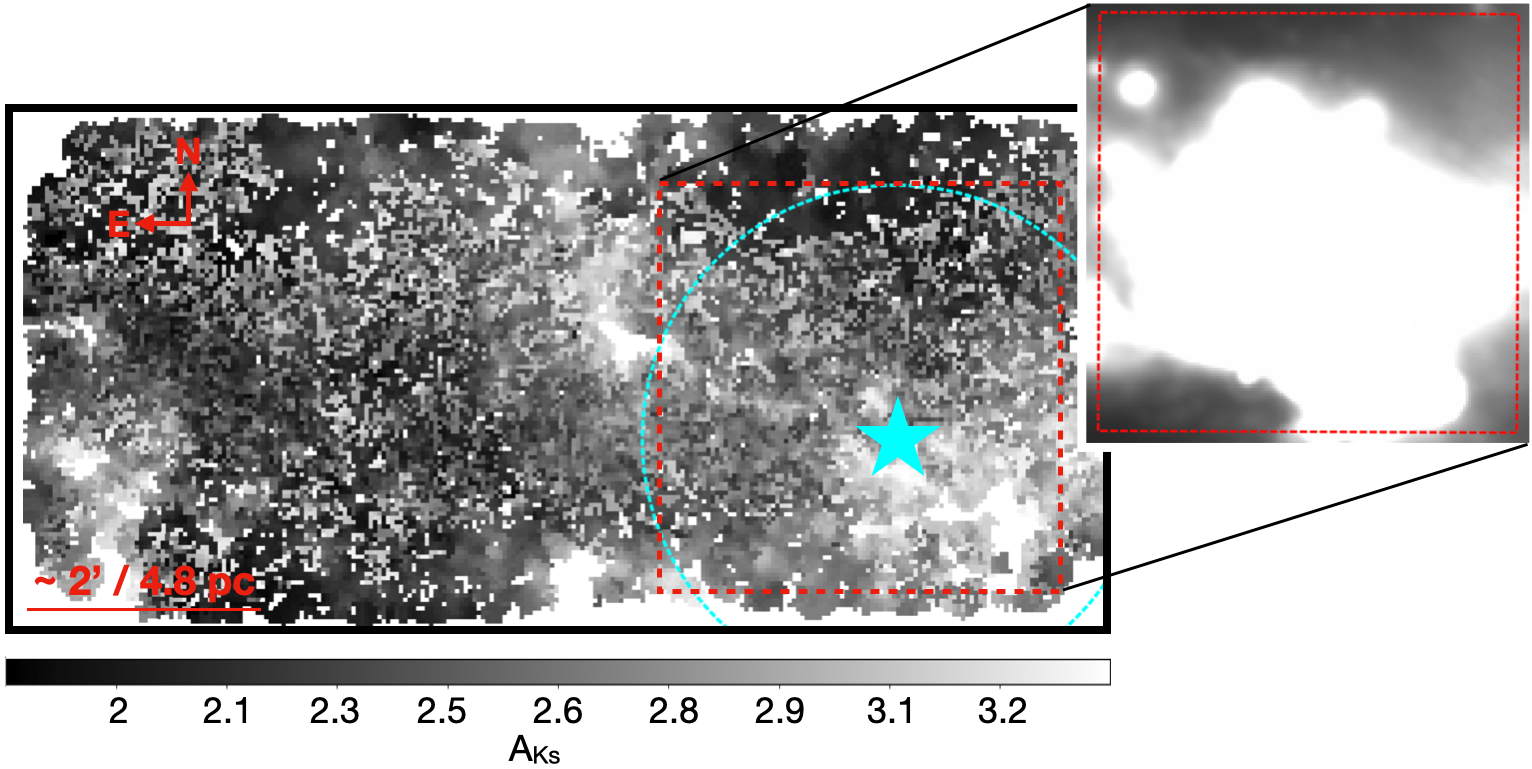}
   \caption{$A_{K_s}$ extinction map for Sgr\,C. The red dashed rectangle indicates a region of particular interest in Sgr\,C (see Sect.\,\ref{small}). The zoomed-in image shows a region of intense hot dust emission (24 microns, \citealt{Carey:2009aa}). The cyan star and dashed circle indicate the centre and approximate extension of the HII region in the Sgr\,C complex \citep{Lang:2010aa}, respectively. The compass corresponds to Galactic coordinates.}
   \label{ext}
    \end{figure}

To deredden the stars belonging to the Sgr\,C region and the control field, we created extinction maps following the methodology described in \citet{Nogueras-Lara:2022ua}. We defined a pixel size of $\sim2''$ and estimated the extinction using RC and red giant stars with similar intrinsic colours to a reference \citep[see e.g.][]{Nogueras-Lara:2021wj}. We computed the extinction value ($A_{K_s}$) for each pixel by applying the equation:

\begin{equation}
A_{K_s} = \frac{H-K_s-(H-K_s)_0}{A_H/A_{K_s}-1}\hspace{0.5cm}
,\end{equation}

\noindent where $A_H/A_{K_s} = 1.84\pm0.03$ \citep{Nogueras-Lara:2020aa}, and the intrinsic colour of reference stars is $(H-K_s)_0=0.10\pm0.01$\,mag \citep{Nogueras-Lara:2021wj}. We also applied an inverse distance weight method considering the five closest stars to a given pixel to account for the distance of the reference stars to each pixel. If fewer than five reference stars were present within a radius of $\sim7.5''$, we did not assign an extinction value for that particular pixel. Additionally, we required that the colour of the reference stars lie within 0.3\,mag of the closest reference star to a given pixel. In this way, we accounted for potential extinction variation along the line of sight \citep[e.g.][]{Nogueras-Lara:2019ad,Nogueras-Lara:2022ua}. Figure\,\ref{ext} shows the obtained extinction map for Sgr\,C with a mean extinction of $A_{K_s}=2.56\pm0.36$\,mag. Somewhat larger extinction values were measured for regions dominated by dust emission (inset in Fig.\ref{ext}). A similar value was obtained for the control region with $A_{K_s}=2.26\pm0.40$\,mag. The uncertainty refers to the standard deviation of the pixel values in each map.

\subsection{Luminosity functions} 
 
We used the above extinction maps to correct the $K_s$ photometry of both the Sgr\,C region and the control field. To prevent the inclusion of over-dereddened stars in the $K_s$ luminosity function, we excluded stars whose dereddened $H-K_s$ colour was more than 2\,$\sigma$ bluer than the mean value of the dereddened distribution of the RC features \citep[e.g.][]{Nogueras-Lara:2019ad,Nogueras-Lara:2022ua}. We built the $K_s$ luminosity function using the `auto' option in the Python function numpy.histogram \citep{Harris:2020aa} to choose the bin width that maximises the Freedman--Diaconis \citep{Freedman1981} and Sturges \citep{doi:10.1080/01621459.1926.10502161} estimators. The associated uncertainty was calculated as the square root of the number of stars per magnitude bin.

Stellar crowding in the Galactic centre significantly reduces the number of stars in the faint end of the obtained luminosity functions \citep[e.g.][]{Nogueras-Lara:2019ad,Schodel:2023aa}. We therefore applied a completeness correction to account for this effect. We divided the Sgr\,C and the control fields into smaller subregions of $\sim2' \times 2'$ and calculated the completeness by estimating the critical distance at which a star is detectable in the vicinity of a brighter star, as explained in \citet{Eisenhauer:1998tg} and \citet{Harayama:2008ph}. The final completeness solution for each field was computed by considering the mean and the standard deviation of the values obtained for each of the subregions. We applied the completeness solution to the $K_s$ luminosity functions for each field, setting a lower limit of $\sim90$\,\% of data completeness. This threshold corresponds to the value at which the Sgr\,C $K_s$ luminosity function is limited by sensitivity (i.e. the number of stars in the luminosity function drastically drops beyond the adopted limit).

\subsection{Fit of the $K_s$ luminosity functions}

To derive the stellar population in the Sgr\,C and the control regions, we fitted the $K_s$ luminosity functions with a linear combination of theoretical models, as explained in \citet{Nogueras-Lara:2019ad,Nogueras-Lara:2022ua} and \citet{Schodel:2023aa}. We used Parsec\footnote{generated by CMD 3.6 (http://stev.oapd.inaf.it/cmd)} \citep{Bressan:2012aa,Chen:2014aa,Chen:2015aa,Tang:2014aa,Marigo:2017aa,Pastorelli:2019aa,Pastorelli:2020wz} and MIST models \citep{Paxton:2013aa,Dotter:2016aa,Choi:2016aa} with twice solar metallicity \citep[e.g.][]{Schultheis:2019aa,Nogueras-Lara:2019ad,Schultheis:2021wf,Nogueras-Lara:2022ua}, and similar stellar ages (14, 11, 8, 6, 3, 1.5, 0.6, 0.4, 0.2, 0.1, 0.04, 0.02, 0.01, and 0.005\,Gyr). The ages of the models were chosen to account for the relevant changes in the shape of the $K_s$ luminosity function with age, and also to specifically investigate the potential presence of the young stars with a good resolution \citep[e.g.][]{Nogueras-Lara:2022ua}. 


We restricted the bright end of the luminosity function to $K_s>8$\,mag to avoid saturation problems that are also present in the SIRIUS IRSF data that we used to correct the GALACTICNUCLEUS photometry \citep[see Sect.\,saturation and supplementary Fig.\,4 in][]{Nogueras-Lara:2022ua}. Given the lower extinction in the control field, we limited the bright end of its luminosity function to $K_s = 8.5$\,mag, as done in previous work \citep{Nogueras-Lara:2022ua}. We included a Gaussian smoothing factor to account for potential different distances between the stars \citep[the NSD scale length is $\sim100$\,pc][]{gallego-cano2019,Sormani:2022wv} and some residual reddening, as well as a parameter to consider the distance towards the stellar populations.

\begin{figure}
   \includegraphics[width=\linewidth]{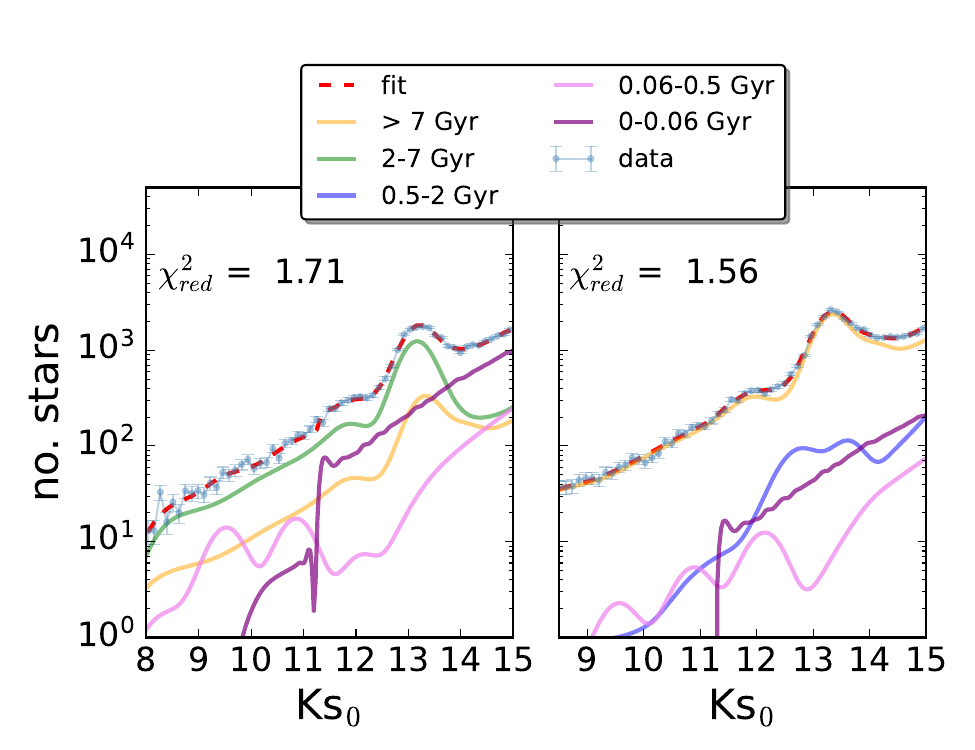}
   \caption{$K_s$ luminosity functions obtained for Sgr\,C (left panel) and the control field (right panel). The Parsec model fit and the reduced $\chi^2$ are indicated in each panel. $Ks_0$ refers to the dereddened $K_s$ magnitude.}
   \label{KLFs}
    \end{figure}

Figure\,\ref{KLFs} shows the luminosity functions and the corresponding fits obtained by applying a $\chi^2$ minimisation when using Parsec models. To obtain the final results, we ran Monte Carlo simulations and created 1000 samples from the original luminosity function by randomly varying the number of stars per bin, assuming their corresponding uncertainty as the standard deviation of the distribution \citep{Nogueras-Lara:2022ua}. We fitted each of the 1000 samples with the Parsec and MIST models and averaged over the results. We combined the models with similar ages into five age bins, merging models with similar ages to decrease the degeneracies ($>7$, 2-7, 0.5-2, 0.06-0.5, and 0-0.06\,Gyr). We obtained the contribution of each age bin to the stellar population by combining the results for Parsec and MIST models, quadratically propagating the corresponding uncertainties. 

\begin{figure}
   \includegraphics[width=\linewidth]{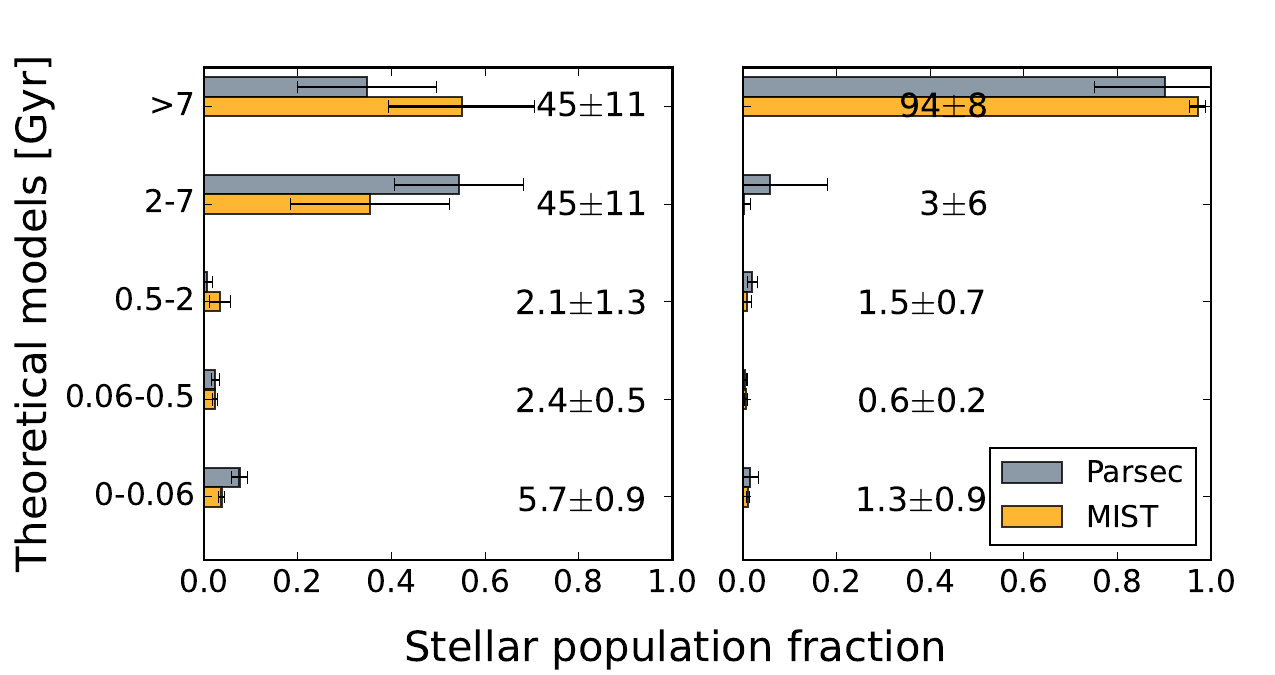}
   \caption{Stellar populations in Sgr\,C (left panel) and the control field (right panel). The numbers in each panel indicate the percentage contribution of each age bin and its associated uncertainty.}

   \label{res}
    \end{figure}

Figure\,\ref{res} shows the results. We find the stellar populations in Sgr\,C and the control field to be significantly different. Sgr\,C shows a prominent contribution from young stars that is around six times larger than that of the control field. Moreover, approximately 50\,\% of the total stellar mass in the Sgr\,C region is attributable to an intermediate-age stellar population (2-7\,Gyr), which is not present in the control field (totally dominated by old stars).

\subsection{Total mass of young stars}

We computed the total mass initially formed in the analysed Sgr\,C region using Parsec models that are calibrated in mass \citep[e.g.][]{Bressan:2012aa,Chen:2014aa,Chen:2015aa,Tang:2014aa,Marigo:2017aa,Pastorelli:2019aa,Pastorelli:2020wz}. We adjusted the bin width of the $K_s$ luminosity function to match that of the theoretical models and calculated the stellar mass by combining the contribution of each of the 14 models. We obtained a total stellar mass of $(8.1\pm0.9)\cdot 10^6$\,$M_\odot$, where the values were estimated using the mean and the standard deviation of the mass distribution for the 1000 Monte Carlo samples. To estimate the total mass of young stars, we considered that $5.7\pm0.9$\,\% of the total stellar mass was due to stars in the youngest age bin (5, 10, 20, 40\,Myr, see Fig.\,\ref{res}). We obtain that $(4.6\pm 0.1)\cdot10^5$\,$M_\odot$ of stars are younger than 60\,Myr.

\subsection{Systematic uncertainties}

We assessed potential sources of systematic uncertainty on the analysis of the Sgr\,C luminosity function, without finding any significant variation with respect to the obtained results:

1. Initial mass function. Given that our sample consists almost entirely of giant stars, our results are almost completely insensitive to a change in the initial mass function \citep{Schodel:2023aa}. We used a Salpeter function \citep{Salpeter:1955qo} for the MIST models, and a Kroupa one \citep{Kroupa:2013wn} for the Parsec models. We made this different choice for the Parsec models because this is the only available initial mass function that also accounts for unresolved binaries. In any case, we also verified that using a Salpeter initial mass function for the Parsec models does not change our results in any significant way.

2. Bin width of the luminosity function. We repeated the analysis by assuming half and double the bin width previously estimated using the Python function numpy.histogram. 

3. Bright end of the luminosity function. We assumed a bright end of the dereddened $K_s$ luminosity function of 7.75\,mag (0.25\,mag brighter than the original one) and repeated the analysis. 

4. Faint end of the luminosity function. We repeated the analysis considering different limits for the faint end. We assumed a completeness of $\sim93\,\%$ and $\sim88\,\%$, which supposes a difference of $\pm0.4$\,mag in the faint end of the luminosity function. 

5. Extinction map. We created a new extinction map, increasing the radius to search for reference stars to $10''$ and choosing seven reference stars instead of five.

6. Metallicity of the stellar population. We repeated the analysis using Parsec models with solar metallicity.

\section{Discussion}

The high mass of young stars that we detect in Sgr\,C suggests the presence of a young stellar association that might have originally formed several stellar clusters \citep[the upper mass limit for a young cluster in the Galactic centre is $\sim10^4$\,$M_\odot$;][]{Trujillo-Gomez:2019ty}. The fact that there are no obvious stellar over-densities in Sgr\,C indicates that the young stellar population is already dispersed, which allows us to roughly estimate the age of the young stars to be $\gtrsim5-10$\,Myr, which corresponds to the time required to dissolve even massive clusters in the NSD \citep[][]{Portegies-Zwart:2002fk,Kruijssen:2014aa}.

Our analysis reveals that the Sgr\,C stellar population is similar to that in Sgr\,B1 \citep{Nogueras-Lara:2022ua}. In both regions, there is an excess of young stars in comparison to the central area of the NSD. Additionally, they exhibit a similar contribution ($\sim40-50\,\%$ of the total stellar mass) from an intermediate-age stellar population, with ages ranging from 2 to 7\,Gyr. This is also significantly different from the stellar population in the central regions of the NSD, which is dominated by old stars \citep[e.g.][]{Nogueras-Lara:2019ad,Schodel:2023aa}, and indicates the presence of an age gradient with increasing stellar ages towards the centre of the NSD.

\subsection{Young stars in a key Sgr\,C subregion}
\label{small}

To better compare Sgr\,C and Sgr\,B1, we applied our luminosity function technique to a relatively small region of $\sim50$\,pc$^2$ (see Fig.\,\ref{scheme}), which likely contains an even higher fraction of young stars, as was done for a similar region in Sgr\,B \citep{Nogueras-Lara:2022ua}. Figure\,\ref{scheme} shows the chosen region, which includes the two known Wolf Rayet (WCL) stars in Sgr\,C \citep[circles in Fig.\,\ref{scheme}; for further details see][]{Clark:2021up}, and several $H_2CO$ and $CH_3OH$ masers tracing recent star formation \citep{Caswell:1996aa,Lu:2019aa}. This area is also characterised by strong 24\,micron emission \citep{Carey:2009aa}, revealing the presence of hot dust likely associated to the HII region in Sgr\,C (see Fig.\,\ref{ext}).

\begin{figure}
   \includegraphics[width=\linewidth]{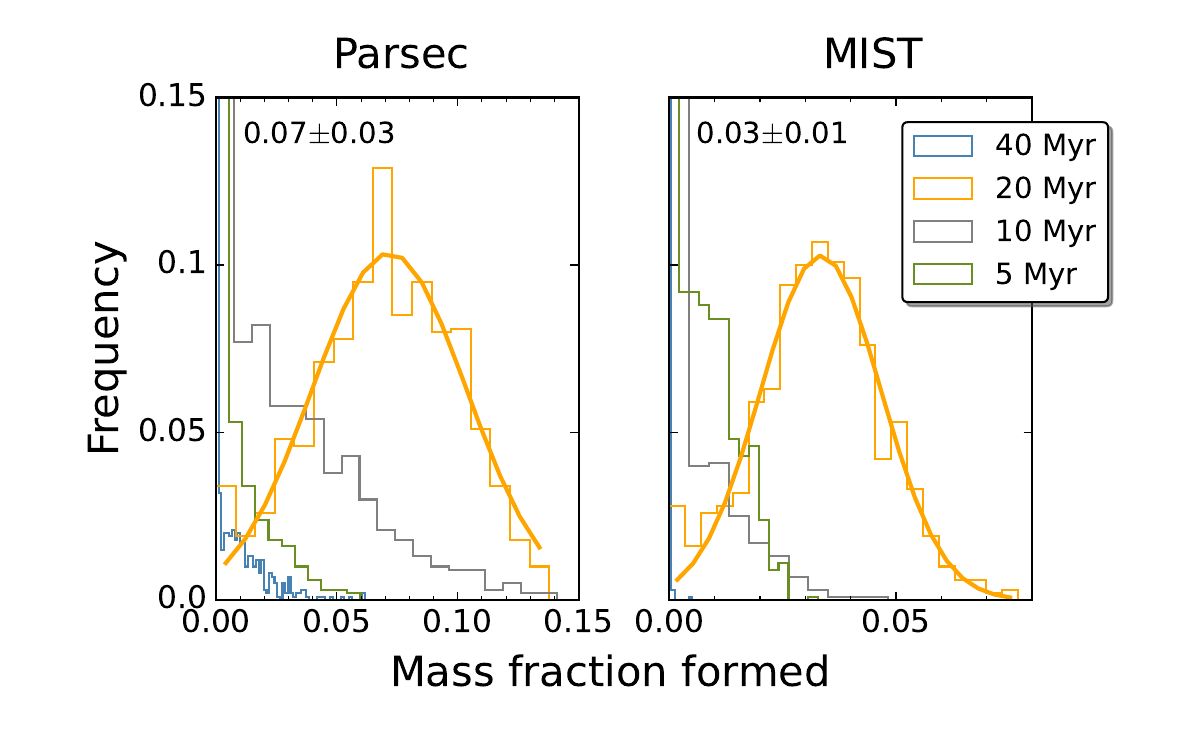}
   \caption{Contribution of the young stellar models to the $K_s$ luminosity function fit in the region dominated by hot dust emission in Sgr\,C (see Fig.\,\ref{scheme}). The solid line shows a Gaussian fit to the contribution of the 20\,Myr model for Parsec (left panel) and MIST (right panel) models. The mean and the standard deviation of the Gaussian fits are indicated in each panel.}

   \label{Young_MC}
    \end{figure}

We find that the young stars (age bin <60\,Myr) in this region account for $\sim7$\,\% of the total stellar mass, implying the presence of $\sim1.5\cdot10^5$\,$M_\odot$ of young stars. We also estimated the age of the stellar population by computing the contribution of the young models (5, 10, 20, 40\,Myr) to each of the Monte Carlo samples. Figure\,\ref{Young_MC} shows the obtained results. We find that the 20\,Myr model is present in $\gtrsim98\,\%$ of the 1000 Monte Carlo samples, and accounts for $\sim5$\,\% of the total stellar mass when averaging over the results from Parsec and MIST models. This enables us to estimate that approximately 70\,\% of the stellar mass attributed to young stars in this region is around 20 million years old.

Assuming a circular velocity for the NSD of $\sim100$\,km/s \citep[e.g.][]{2015ApJ...812L..21S,Sormani:2022wv} and a distance of $\sim70$\,pc for Sgr\,C from the supermassive black hole, we estimate the rotation period of the Sgr\,C stellar population to be $\sim4$\,Myr. This implies that the detected young stellar population had enough time to complete several orbits around the NSD and did not  form in situ. The presence of very young stars in the region, such as the $H_2CO$ and $CH_3OH$ masers \citep{Caswell:1996aa,Lu:2019aa}, could be due to ongoing star formation triggered by the outflows from the ionising stars of the detected $\sim20$\,Myr stellar population. This mechanism might also explain the presence of the HII emission, in a similar way as proposed for the Sgr\,B1 region \citep[e.g.][]{Simpson:2018aa}. Moreover, Sgr\,C has been tentatively proposed as a connection point with one of the gas and dust streams linking the central molecular zone and the NSD to the Galactic bar \citep[e.g.][]{Molinari:2011fk,Henshaw:2022vl}. This would explain the presence of a significant amount of dust, which is heated by the ionising stars, causing the intense 24\,micron emission \citep{Carey:2009aa}.

An analogous analysis carried out on a $\sim40$\,pc$^2$ region in Sgr\,B1 revealed the presence of a similar stellar mass of a slightly younger stellar population ($\sim5-10$\,Myr). Therefore, Sgr\,C likely represents the future state of the young stellar association in Sgr\,B1, and may provide crucial information for understanding the evolution of young stars in the Galactic centre and addressing the missing clusters problem.

\subsection{Presence of an intermediate-age stellar population}

A study of the external Milky Way-like galaxies from the TIMER survey \citep{Gadotti:2019aa} suggested that NSDs form inside-out from gas funnelled by galactic bars towards the innermost regions of the galaxies \citep{Bittner:2020aa}. The recent analysis of the NSD stellar population along the line of sight revealed that the innermost region of the NSD is dominated by old stars, whereas the outer edge exhibits, on average, a younger stellar population \citep{Nogueras-Lara:2023aa}. This finding supports the inside-out formation scenario and indicates that the NSD probably originated from gas funnelling towards the Galactic centre  through the Galactic bar.

\begin{figure}
    \begin{center}
   \includegraphics[width=0.7\linewidth]{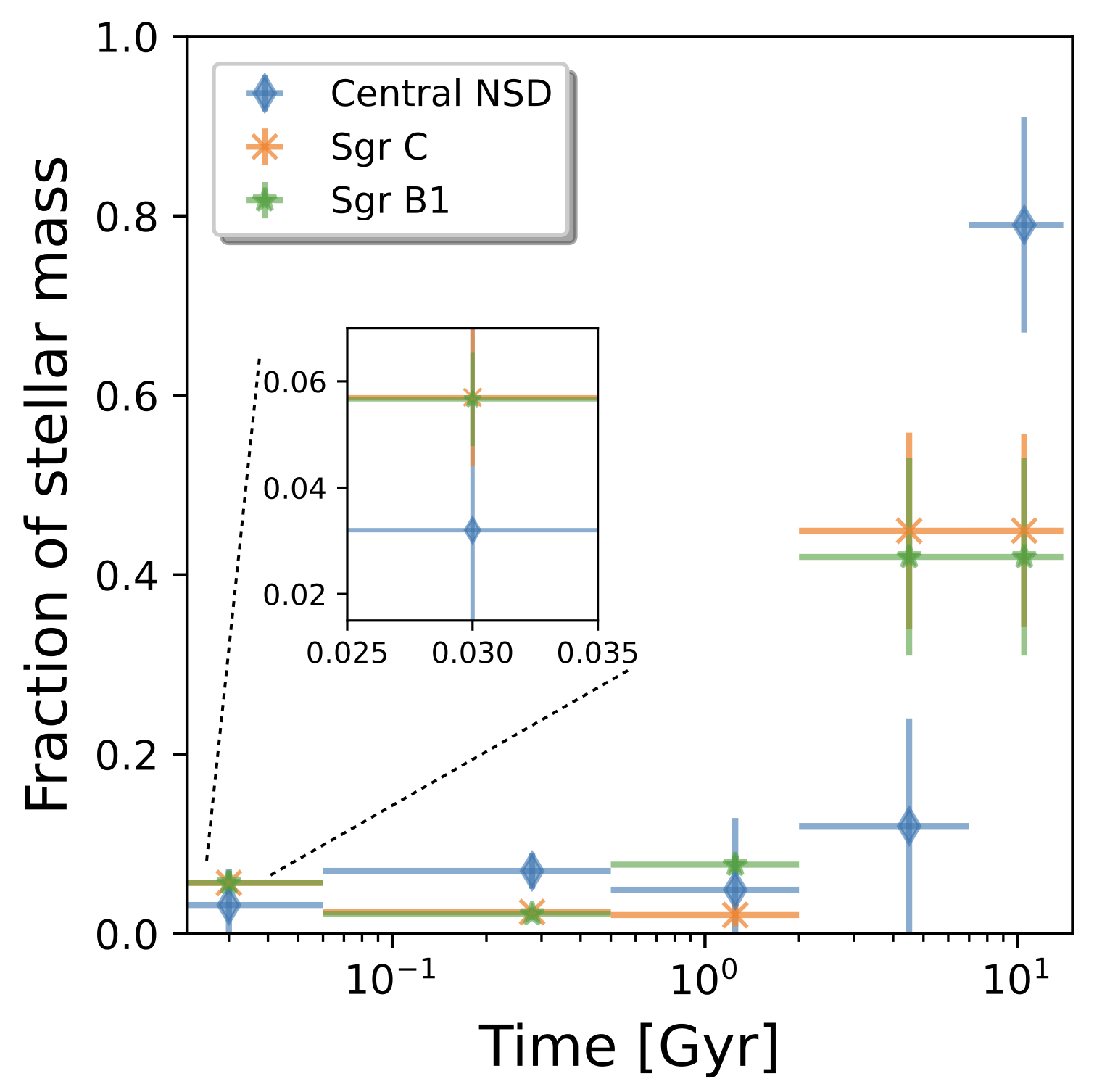}
   \caption{Comparison between the stellar population present in the central region of the NSD, Sgr\,B1  \citep[data extracted from Fig.\,4 in][]{Nogueras-Lara:2022ua}, and Sgr\,C (this work). The zoomed-in image shows the youngest age bin, where Sgr\,B1 and Sgr\,C show an excess of young stars in comparison to the central region of the NSD.}

   \label{comparison}
   \end{center}
    \end{figure}

The study presented in this letter, along with the results on the Sgr\,B1 region \citep{Nogueras-Lara:2022ua}, allowed us to assess the inside-out formation growth of the NSD by analysing the stellar population at its centre \citep{Nogueras-Lara:2019ad,Schodel:2023aa} and its edges (Sgr\,B1 and Sgr\,C, see Fig.\,\ref{scheme}). Figure\,\ref{comparison} shows our findings. While the central region of the NSD is dominated by an old stellar population ($\gtrsim80$\,\% of the total stellar mass is older than $\sim7$\,Gyr), Sgr\,C shows a stellar population akin to Sgr\,B1 (see Fig.\,\ref{comparison}). We find that $\sim50\,\%$ of its stellar mass  falls within the age range of 2-7\,Gyr. This supports the inside-out formation scenario and agrees with previous findings on the stellar population along the NSD line of sight \citep{Nogueras-Lara:2023aa}.

Our results slightly disagree with the recent analysis of Mira variables in the NSD \citep{Sanders:2023aa}, which only reveals weak evidence of inside-out growth. Nevertheless, these latter authors report a noteworthy population of Mira variables with periods of $\sim400$\,days (equivalent to an age of $\sim5-6$\,Gyr), which possibly correspond to stars accounting for the age gradient that we report here. \citet{Sanders:2023aa} acknowledged that this stellar population was not distinctly identified in their modelling, which is possibly because of the smoothing applied to their spline model and the presence of particularly large uncertainties.

\section{Conclusion}

In this letter, we present an analysis of the stellar population in Sgr\,C situated at the western edge of the NSD. Our findings reveal the presence of several $10^5\,M_\odot$ of young stars, comprising $\sim6\,\%$ of the stellar mass in the region. This fraction is roughly six times larger than what we observed in a control field in the central region of the NSD, evidencing a clear excess of young stars. To best of our knowledge, our results constitute the first detection of a large population of young stars at the western edge of the NSD. This shows that the asymmetric gas distribution in the central molecular zone \citep[e.g.][]{Bally:1988vf,Sormani:2018aa} does not necessary imply an asymmetric distribution of young stars in the NSD. 

Moreover, we compared the young stars in Sgr\,C with those in Sgr\,B1 located at the eastern edge of the NSD \citep{Nogueras-Lara:2022ua}, and conclude that those in Sgr\,C are likely older. This implies that Sgr\,C may provide insights into the future evolution of the young stellar population in Sgr\,B1. In this way, it is now possible to identify regions that sample the various stages of young stellar evolution in the Galactic centre. Sgr\,B2 is a region of ongoing star formation \citep[e.g.][]{Ginsburg:2018uq}, the young Arches and Quintuplet clusters show a stellar population with $\sim$2-5\,Myr \citep[e.g.][]{Najarro:2004kx,Martins:2008aa,Clark:2018aa,Clark:2018ab}, and finally Sgr\,B1 and Sgr\,C contain stellar associations with $\sim10$ and 20\,Myr, respectively.

We also find that $\sim50$\,\% of the total stellar mass in Sgr\,C is attributed to intermediate-age stars ($\sim2-7$\,Gyr). We compared our results with the central regions of the NSD and Sgr\,B1 and conclude that they support the presence of an age gradient in the NSD. Old stars dominate its central regions \cite[e.g.][]{Nogueras-Lara:2019ad,Schodel:2023aa}, while a younger stellar population (on average) appears close to the edge of the NSD \citep{Nogueras-Lara:2022ua,Nogueras-Lara:2023aa}. This points towards an inside-out formation of the NSD and suggests that bar-driven processes found in external barred galaxies are also at play in the Milky Way \citep{Bittner:2020aa}.

  \begin{acknowledgements}
We thank the anonymous referee for helpful comments and suggestions that improved this manuscript. FN-L gratefully acknowledges the sponsorship provided by the European Southern Observatory through a research fellowship. This work is based on observations made with ESO Telescopes at the La Silla Paranal Observatory under program ID195.B-0283. We thank the staff of ESO for their great efforts and helpfulness.

\end{acknowledgements}

\bibliography{BibGC.bib}
\end{document}